\documentclass[twocolumn,           % Format : preprint, twocolumn
               showpacs,            % Pacs : showpacs, noshowpacs
               preprintnumbers,     % Preprint: preprintnumbers,
               aps,                 % Society: ...
               prd,                 % Journal Style : pra, prb, prc, prd, pre,
               a4paper,             % Size : a4paper, ...
               superscriptaddress,  % Affiliation (Title) : groupedaddress,
               nofootinbib,         % Footnote: footinbib, nofootinbib
               tightenlines,        % Remove additional spaces in a line
               floats,floatfix      % Floating pictures and tables
               ]{revtex4}

\usepackage{graphicx}
\usepackage{subfigure}
\usepackage{latexsym}
\usepackage{amsmath,amssymb}        % amssymb includes amsfonts
\usepackage[draft=false]{hyperref}

\input colordvi

\begin{document}

%======================================%
%<<<<<<<<<<<< TITLE PAGE >>>>>>>>>>>>>>%
%======================================%

\title{Present and future evidence for evolving dark energy}
\author{Andrew R.~Liddle}
\affiliation{Astronomy Centre, University of Sussex, Brighton BN1 9QH,
United Kingdom}
\author{Pia Mukherjee}
\affiliation{Astronomy Centre, University of Sussex, Brighton BN1 9QH,
United Kingdom}
\author{David Parkinson}
\affiliation{Astronomy Centre, University of Sussex, Brighton BN1 9QH,
United Kingdom}
\author{Yun Wang}
\affiliation{Department of Physics and Astronomy, University of
  Oklahoma, Norman, OK 73019, USA} 
\date{\today}
\pacs{98.80.-k \hfill astro-ph/0610126}
\preprint{astro-ph/0610126}

%======================================%
%<<<<<<<<<<<<< ABSTRACT >>>>>>>>>>>>>>>%
%======================================%

\begin{abstract}
We compute the Bayesian evidences for one- and two-parameter models of
evolving dark energy, and compare them to the evidence for a
cosmological constant, using current data from Type Ia supernova,
baryon acoustic oscillations, and the cosmic microwave background.  We
use only distance information, ignoring dark energy perturbations.  We
find that, under various priors on the dark energy parameters,
$\Lambda$CDM is currently favoured as compared to the dark energy
models. We consider the parameter constraints that arise under
Bayesian model averaging, and discuss the implication of our results
for future dark energy projects seeking to detect dark energy
evolution. The model selection approach complements and extends the
figure-of-merit approach of the Dark Energy Task Force in assessing
future experiments, and suggests a significantly-modified
interpretation of that statistic.
\end{abstract}

\maketitle

%======================================%
%<<<<<<<<<<<<<< ARTICLE >>>>>>>>>>>>>>>%
%======================================%

\section{Introduction}

A key challenge for cosmology is to uncover the nature of the force
which is causing the Universe to expand at an accelerating rate
today. The cause, dubbed dark energy, could be an unknown energy
component with negative pressure \cite{Peebles88}, a
modification of general relativity \cite{DGP00}, or simply a
cosmological constant. For reviews on the subject, see for example
Ref.~\cite{revs}.

There are many planned and proposed dark energy experiments that aim
to constrain dark energy parameters, using a combination of
complementary techniques. These include the luminosity
distance--redshift relation of Type Ia supernovae (SNe Ia), the
angular-diameter distance--redshift and expansion rate--redshift
relations measured by baryon acoustic oscillations (BAO), and use of
weak gravitational lensing to probe the growth rate of structures. The
cosmic microwave background (CMB) also provides a very useful handle
on dark energy by pinning down the distance to the last-scattering
surface, and also via the Integrated Sachs--Wolfe effect and by
detecting clusters through the Sunyaev--Zel'dovich effect. Approaches
to constraining dark energy were overviewed in the recent report of
the DoE/NASA/NSF Dark Energy Task Force (DETF) \cite{DETF}.

A primary aim of future experiments is to distinguish evolving dark
energy from a cosmological constant.  When seeking to compare models,
especially with different numbers of variable parameters, one should
use the concepts of model selection rather than those of parameter
estimation (e.g.~Refs.~\cite{Trotta05,AG}). Model selection quantifies
how well the data conform to the overall predictions of a model, which
depends on model dimensionality and model priors. In addressing the
primary goal, a satisfactory representation of many evolving dark
energy models turns out to be an unknown energy component with
equation of state $w(a)=w_0+w_a(1-a)$, where $a$ is the scale
factor. Simpler alternatives may be the constant $w$ model with
negative pressure, and the cosmological constant model with fixed
$w=-1$. This is a natural area for the application of model selection
statistics \cite{MPL05,M06}, which we take up in this paper. Here we
update and extend work by Saini et al.~\cite{Saini04}, who were the
first to apply Bayesian model selection to dark energy models. For
alternative views on determining the number of dark energy parameters,
see Ref.~\cite{BCK}.

We do not consider growth-of-structure constraints, which ultimately
will be required to distinguish between dark energy and modified
gravity models for the acceleration \cite{DETF,growth}. In the
phenomenological approach adopted here, the dynamical evolution of $w$
could be attributed to either phenomenon. At present, the structure
formation growth factor theory is known only for specific modified
gravity models, and further development is needed before such models
can be usefully considered in the model selection framework. In any
case, at the present time these observations are not competitive with
the ones we use.

In this paper we compute the Bayesian evidence for evolving dark
energy versus that of a cosmological constant given current distance
measurements from CMB, SN Ia, and BAO data, ignoring dark energy
perturbations. In light of this result, we discuss the probability
that future experiments will detect evolving dark energy, and the
implications of this in assessing the capabilities of future
experiments.

\section{Method and models}

Bayesian model selection extends the usual parameter estimation
framework by assigning probabilities to \emph{sets} of parameters,
known as models, as well as the usual probability distributions of
parameter values for each specific choice of model.  The key statistic
of Bayesian model selection is the \emph{Bayesian evidence} $E$, being
the average likelihood of the model over its prior parameter ranges
\cite{MacKay,Gregory}. This quantity updates the prior model
probability to the posterior model probability, enabling one to
compare different models according to their probability.

The use of the Bayesian evidence has lagged behind parameter
estimation techniques in the cosmology literature because of the
difficulty in computing the required integral to high accuracy, so as
to be able to distinguish between the models of interest.  The nested
sampling algorithm, proposed by Skilling \cite{Skilling} and
implemented for cosmological applications by some of us in
Ref.~\cite{MPL05}, has proven to be computationally efficient and
accurate. It is a simple algorithm and more general than thermal
methods as used in Refs.~\cite{Beltran05,Bridges06}. For instance,
nested sampling can handle multiphase problems, in which $\ln L$ is
not a concave function of $\ln X$, where $X$ is the cumulative
probability mass within isolikelihood surfaces and $L$ the likelihoods
of the surfaces --- thermal methods fail on such problems (John
Skilling, private communication).

We use the nested sampling algorithm to compute the Bayesian evidences
\cite{Skilling, MPL05,PML06}. Our code, called CosmoNest, is available
at the URL www.cosmonest.org. As compared to the public version, it
was modified so that instead of using the power spectra for each
model, it used the data and likelihoods described in the next
section. As we are not computing power spectra the calculation
proceeds very swiftly, taking just a few minutes to obtain multiple
estimates of the evidence of a model. The estimates can then be
combined into a mean evidence and an error on that mean.

We consider five different models in all, corresponding to different
parametrizations of the equation of state $w$ and/or different
parameter priors. The basic models are $\Lambda$CDM ($w=-1$, Model I),
a one-parameter model with constant $w$, and a two-parameter model
$w(a)=w_0+w_a(1-a)$, where $w_0$ and $w_a$ are constants. This last
parametrization, introduced by Chevallier and Polarski \cite{ChevPol},
is a good approximation to many dark energy models, while the constant
$w$ model is purely phenomenological. In addition to the equation of
state, each model requires two further parameters to complete its
specification, the matter density $\Omega_{{\rm m}}$ and the Hubble
constant $H_0$.

For the latter two parametrizations, we make two separate choices of
prior in order to explore this dependence. For the constant $w$ case
these are $-1 \leq w \leq -0.33$ (Model II) and $-2 \leq w \leq -0.33$
(Model III), the former enforcing the weak energy condition and the
latter allowing phantom models. For the two-parameter model, Model IV
has flat priors of $-2 \leq w_0 \leq -0.33$, $-1.33 \leq w_a \leq
1.33$ (the prior on $w_a$ being particularly arbitrary), while Model V
corresponds to the quintessence prior of $-1 \leq w(a) \leq 1$ imposed
between $z=0$ and $2$.

\section{Observational data}

We use data in a manner very similar to Wang and Mukherjee
\cite{WM06}, which can be consulted for more details.

We use the CMB shift parameter measured by the three-year Wilkinson
Microwave Anisotropy Probe (WMAP) observations
\cite{Hinshaw06,Spergel06}, of $R=1.70 \pm 0.03$ \cite{WM06}, which is
mostly independent of assumptions made about dark energy. The shift
parameter $R$ is \cite{Bond97}
\begin{equation}
R\equiv \Omega_{{\rm m}}^{1/2}\int_0^{z_{{\rm CMB}}} \frac{dz'}{E(z')}
\,, 
\end{equation}
where $z_{{\rm CMB}}$ is the redshift of recombination. In a flat
Universe
\begin{equation}
E(z)\equiv \left[\Omega_{{\rm m}} (1+z)^3 + (1-\Omega_{{\rm
m}})\frac{\rho_X(z)}{\rho_X(0)} \right]^{1/2}\,,
\end{equation}
with $\rho_X$ denoting the dark energy density given by
\begin{equation}
\frac{\rho_X(z)}{\rho_X(0)}=\exp \left\{ \int_0^z dz'
\frac{3[1+w(z')]}{1+z'} \right\}\,.
\end{equation} 
In that case $R= \left({\Omega_{{\rm m}} H_0^2}\right)^{1/2} r(z_{{\rm
CMB}})/c$, and is well determined as both $\Omega_{{\rm m}} h^2$ and
$r(z_{{\rm CMB}})$ are accurately measured by CMB data.

We use the BAO measurement from the Sloan Digital Sky Survey (SDSS)
luminous red galaxies, $d_V(0.35)=1.300 \pm0.088 \rm Gpc$
\cite{Tegmark06}, obtained from power spectrum estimates and
consistent with the result of Ref.~\cite{Eisen05} obtained using the
estimated correlation function. Here the distance parameter is
\begin{equation}
d_V(z_{{\rm BAO}})=\left[ r^2(z_{{\rm BAO}})\, \frac{cz_{{\rm
BAO}}}{H(z_{{\rm BAO}})} \right]^{1/3}\,,
\end{equation}
where $r(z)$ is the comoving distance, and $H(z)$ is the Hubble
parameter. For the SDSS luminous red galaxies, the mean survey
redshift is $z_{{\rm BAO}}=0.35$.\footnote{The SDSS BAO result has
been computed for a scalar spectral index value of $n_{{\rm S}}=0.98$,
and should be scaled by $(n_{{\rm S}}/0.98)^{-0.35}$ \cite{Eisen05}
for a different `best-fit' $n_{{\rm S}}$. For $n_{{\rm S}}\simeq 0.95$
following WMAP3 \cite{Spergel06} this is an insignificant factor,
which however we do include.}

We use SN Ia data from the HST/GOODS programme \cite{Riess04}
(Riess04) and the first year Supernova Legacy Survey \cite{Astier05}
(Astier05), together with nearby SN Ia data. The comparison of results
from these two SN Ia datasets provides a consistency check. We do not
combine the two SN Ia datasets, as they have systematic differences
in data processing; see the discussion in Ref.~\cite{WM06}.

We use the Riess04 `gold' sample flux-averaged with $\Delta
z=0.05$. This sample includes 9 SNe Ia at $z>1$, and appears to have
systematic effects from weak lensing, or another effect that mimics
weak lensing qualitatively. This would bias the distance estimates
somewhat without flux averaging \cite{WangTegmark04,Wang05}, and so
we use it on these SNe \cite{WangMukherjee04}.

\begin{table*}
\caption{\label{t1} The mean $\Delta \, \ln E$ relative to the
    $\Lambda$CDM model together with its uncertainty, the information
    content $H$, the minimum $\chi^2$, and the parameter constraints,
    for each of the models considered and for each of two data
    combinations.  Uncertainties on $H_0$ are statistical only, and do
    not include systematic uncertainties.  The models differ by virtue
    of the number of free parameters, here in the dark energy sector,
    and/or the priors on those parameters. For reference, $\ln E$ for
    the $\Lambda$CDM model was found to be $-20.1\pm0.1$ for the
    compilation with Riess04 and $-52.3\pm0.1$ for that with
    Astier05.}
\begin{center}
\begin{tabular}{|c|cccc|}
\hline
data used  & \multicolumn{4}{c|}{Model} \\
\hline
WMAP+SDSS+ & $\Delta \ln E$ & $H$ & $\chi^2_{{\rm min}}$ & parameter
constraints \\ 
\hline
\hline
  & \multicolumn{4}{c|}{Model I: $\Lambda$} \\
\hline
{\rm Riess04} & $0.0$ & 5.7 & 30.5 &
$\Omega_{{\rm m}}=0.26\pm0.03$, $H_0=65.5\pm1.0$\\ 
{\rm Astier05} & $0.0$ & 6.5 & 94.5 &
$\Omega_{{\rm m}}=0.25\pm0.03$, $H_0=70.3\pm1.0$\\ 
\hline
\hline
  & \multicolumn{4}{c|}{Model II: constant $w$, flat prior $-1 \leq
  w\leq -0.33$}\\  
\hline
{\rm Riess04} & $-0.1\pm0.1$ & 6.4 & 28.6 &
$\Omega_{{\rm m}}=0.27\pm0.04$, $H_0=64.0\pm 1.4$,
$w<-0.81,-0.70$\footnotemark[1] \\ 
{\rm Astier05} & $-1.3\pm0.1$ & 8.0 & 93.3 &
$\Omega_{{\rm m}}=0.24\pm0.03$, $H_0=69.8\pm1.0$,
$w<-0.90,-0.83$\footnotemark[1] \\ 
\hline
\hline
  & \multicolumn{4}{c|}{Model III: constant $w$, flat prior $-2 \leq w
  \leq-0.33$}\\ 
\hline
{\rm Riess04} & $-1.0\pm0.1$ & 7.3 & 28.6 &
$\Omega_{{\rm m}}=0.27\pm0.04$, $H_0=64.0\pm 1.5$,
$w=-0.87\pm0.1$ \\ 
{\rm Astier05} & $-1.8\pm0.1$ & 8.2 & 93.3 &
$\Omega_{{\rm m}}=0.25\pm0.03$, $H_0=70.0\pm1.0$,
$w=-0.96\pm0.08$ \\ 
\hline
\hline
  & \multicolumn{4}{c|}{Model IV: $w_0$--$w_a$, flat prior $-2 \leq
  w_0 \leq -0.33$, $-1.33 \leq w_a \leq 1.33$}\\   
\hline 
{\rm Riess04} & $-1.1\pm0.1$ & 7.2 & 28.5 &
$\Omega_{{\rm m}}=0.27\pm0.04$, $H_0=64.1\pm1.5$, $w_0=-0.83\pm0.20$,
$w_a=--$\footnotemark[2] \\ 
{\rm Astier05} & $-2.0\pm0.1$ & 8.2 & 93.3 &
$\Omega_{{\rm m}}=0.25\pm0.03$, $H_0=70.0\pm1.0$, $w_0=-0.97\pm0.18$,
$w_a=--$\footnotemark[2] \\ 
\hline
\hline
  & \multicolumn{4}{c|}{Model V: $w_0$--$w_a$, $-1 \leq w(a) \leq 1$
  for $0\leq z\leq 2$}\\
\hline
{\rm Riess04} & $\,$ $-2.4\pm0.1$ $\,$ & $\,\,\,$9.1 $\,\,\,$&
28.5 & $\Omega_{{\rm m}}=0.28\pm0.04$, $H_0=63.6\pm1.3$,
$w_0<-0.78,-0.60$\footnotemark[1], $w_a=-0.07\pm0.34$\\ 
{\rm Astier05} & $-4.1\pm0.1$ & 11.1 & 93.3 &
$\Omega_{{\rm m}}=0.24\pm0.03$, $H_0=69.5\pm1.0$,
$w_0<-0.90,-0.80$\footnotemark[1], $w_a=0.12\pm0.22$ \\ 
\hline
\end{tabular}
\footnotetext[2]{Where constraints on $w$ are shown as upper limits
  only, the values are 68\% and 95\% marginalized confidence limits.}
\footnotetext[3]{$w_a$ is unconstrained in Model IV.}
\end{center}
\end{table*}

We have also added a conservative estimate of the intrinsic dispersion
of SN Ia peak brightness, 0.15 mag, in quadrature with the distance
moduli of Astier05, rather than the smaller intrinsic dispersion
derived by them by requiring a reduced $\chi^2=1$ in their model
fitting. This is because the intrinsic dispersion in SN Ia peak
brightness should be derived from the distribution of nearby SNe Ia,
or SNe Ia from the same small redshift interval if the distribution in
the peak brightness evolves with cosmic time. This distribution
is not well known at present, but will become better known as more SNe
Ia are observed by the nearby SN Ia factory \cite{Aldering02}. By
using the larger intrinsic dispersion, we allow some reasonable margin
for the uncertainties in the SN Ia peak brightness distribution.

\section{Results}

We calculate the Bayesian evidence as our primary model selection
statistic. We also calculate the information content $H$ of the
datasets, the best-fit $\chi^2$ values, and the posterior parameter
distributions within each model. Our main focus is on the evidence and
the parameter distributions.  All of these quantities are by-products
of running CosmoNest to evaluate the evidence of a model~\cite{PML06}.

\subsection{Bayesian evidence $E$}

The interpretational scale introduced by Jeffreys \cite{Jeff} defines
a difference in $\ln E$ of greater than 1 as significant, greater than
2.5 as strong, and greater than 5 as decisive, evidence in favour of
the model with greater evidence.

Our results are summarized in Table~\ref{t1}. The priors on the
equation of state parameters were given earlier and are indicated in
the table. Priors on the additional parameters are $0.1 \leq
\Omega_{{\rm m}}\leq 0.5$ and $40 \leq H_0 \leq 90$. For each model
and data compilation we tabulate $\Delta \, \ln E$, which is the
difference between the mean $\ln E$ of the $\Lambda$CDM model and the
model concerned, plus the error on that difference, obtained from 8
estimates of the evidence of each model.  Thus the $\Lambda$CDM entry
is zero by definition.

We find that the WMAP+SDSS(BAO)+Astier05 data combination 
distinguishes amongst the models more strongly than does 
WMAP+SDSS(BAO)+Riess04 data, while showing the same general
trends. Subsequently, our discussion uses Astier05 throughout. 

\begin{figure*}
\centering
\includegraphics[width=14.5cm]{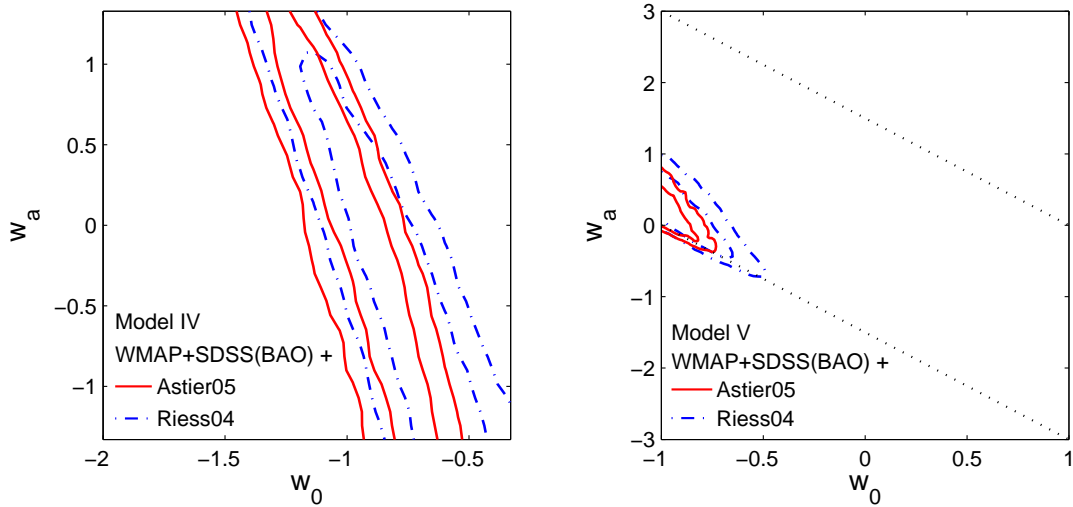}
\caption{68\% and 95\% confidence constraints on $w_0$ and $w_a$ for
model IV (left plot) and model V (right plot). Note the axis ranges
are different, to show the full prior ranges in each case. Model IV
corresponds to a flat prior on the parameters over the range plotted,
and Model V corresponds to a quintessence prior which amounts to a
flat prior within the region shown by dotted lines (the contours go
just a little out of that region due to the effect of binning the
likelihoods of the obtained samples on a grid). The solid contours are
for WMAP+SDSS(BAO)+Astier05, and dot-dashed contours are for
WMAP+SDSS(BAO)+Riess04.}
\label{fig1}
\end{figure*}

Overall, the $\Lambda$CDM model (Model I) is a simple model that
continues to give a good fit to the data. It is therefore rewarded for
its predictiveness with the largest evidence, and remains the favoured
model as found with an earlier dataset (of SNe alone) by Saini et
al.~\cite{Saini04}.  The other models all show smaller evidences,
though none are yet decisively ruled out. Nevertheless, there is
distinct evidence against the two-parameter models, especially from
the compilation including Astier05. Model V has a wider parameter
range than Model IV and fares the worst, receiving a large penalty for
its lack of predictiveness of the data. The one-parameter models lie
somewhere in between.

We interpret these results in the following section.

\subsection{The information $H$}

The information content of the data $H$ is defined as minus
the logarithm of the amount by which the posterior is compressed
inside the prior by the data. We compute it from the posterior
samples generated using nested sampling \cite{PML06}, and
tabulate the values. $H$ gives some indication of how many 
parameters a data set can support, as usually 
$H \approx N\log({\rm signal/noise})$ where $N$ is the number of 
parameters \cite{Skilling}.  If $H$ changes  significantly when new 
parameters are added, then that implies that the data have the 
potential to constrain the additional parameters effectively, and 
therefore have something conclusive to say about the distinction 
between the two models via the evidence. $H$ is similar to the 
effective complexity of a model, as discussed in Ref.~\cite{Kunz06}. 
By definition, $H$ depends on the prior, and the higher $H$ is, the 
better the posterior is confined with respect to the prior.

\begin{figure*}
\centering
\includegraphics[width=14cm]{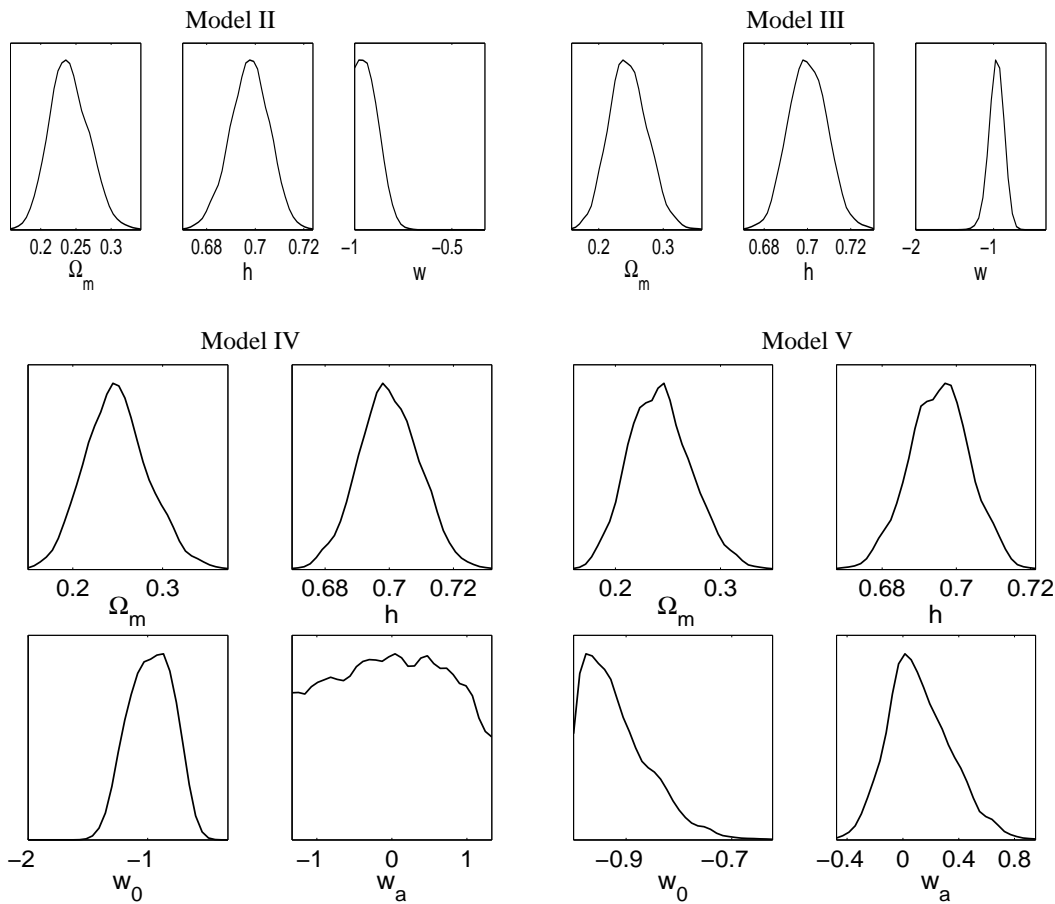}
\caption{Marginalized posterior parameter distributions in Models II,
III, IV, and V, using the WMAP+SDSS(BAO)+Astier05 data combination.}
\label{fig2}
\end{figure*}

\subsection{Best-fit $\chi^2$}

The best-fit $\chi^2$ obtained for each data set is listed in
Table~\ref{t1}, mainly for reference only. They were obtained from the
highest-likelihood point found by the nested sampling algorithm. This
will be close to, though not precisely at, the maximum, because the
stopping criterion for the nested sampling algorithm has to do with
the convergence of the integral that estimates the evidence; the
algorithm is not directly searching for the maximum-likelihood point.
A naive model selection test, formalized as the likelihood ratio test,
compares the difference in these best-fit values to the difference in
number of model parameters. This does not however have a probabilistic
interpretation, as the probability of the model is a property of its
entire parameter range, not simply its best-fit values
\cite{MacKay}. It ignores parameter priors and correlations.

Nevertheless, the lack of any significant improvement in $\chi^2_{{\rm
min}}$ when going from the constant $w$ models to the $w_0$--$w_a$
models could be used to conclude that the dataset is not interested in
going to the two-parameter model.

\subsection{Posterior parameter distributions}

\label{s:params}

Parameter constraints for each of the models, obtained as described in
Ref.~\cite{PML06} from the same samples that were used to compute the
evidence, are tabulated in the final column of
Table~\ref{t1}. Likelihood contours for the dark energy parameters in
Models IV and V are shown in Fig.~\ref{fig1}, the contours in Model V
being significantly cut off by the prior. 1D marginalized parameter
constraints are shown in Fig.~\ref{fig2}.

\section{Discussion: the present picture}

The above results have quantified the impact of current data in
constraining the models we have selected for investigation. There are
considerable modelling uncertainties, both in the choice of parameter
priors for each model, and in assigning prior model probabilities. For
the latter, we have chosen to take them as equal, but anyone who
thinks otherwise can readily account for it; regardless of what
someone thinks about two models before looking at the data (the prior
model probabilities), the evidence unambiguously states how that view
is changed by the data. For the parameter priors, we have analyzed two
choices for each model dimensionality to investigate the extent of the
dependence.

In analyzing data in a situation where the correct choice of model is
unknown, these uncertainties are unavoidable, but one can nevertheless
use the Bayesian framework to draw conclusions. Within the model
selection viewpoint, one should first ask of the status of the various
models under discussion, and only then move on to consider parameter
constraints. 

\subsection{Models}

In comparing the model evidences, there are different ways to
proceed. First, we can consider all five models as independent, so
that converting the $\Delta \, \ln E$ into posterior model
probabilities, assuming equal prior model probabilities, gives 63\%,
17\%, 10\%, 9\%, and 1\% for the five models
respectively. Consequently, while we cannot say that any of the models
is decisively ruled out, the balance of probability is currently
tilted significantly in favour of $\Lambda$CDM, and the two-parameter
equation of state model fares the worst.

\begin{table*}
\caption{\label{t2} Parameter constraints from Bayesian model
averaging using the WMAP+SDSS(BAO)+Astier05 data combination. 
%The pivot parameters derived from the constraints on $w_0$ and $w_a$ 
%are also shown. 
Since the distributions of the dark energy parameters are
generally nongaussian and/or asymmetric about the mean, their 68\% and
95\% marginalized limits are separately indicated. Some confidence
limits for $w_a$ are precisely zero due to the delta-function
contribution from Models I, II, and III superimposed on the extended
tails from Models IV and V.}
\begin{center}
\begin{tabular}{|c|c|}
\hline
models used  & parameter constraints \\
\hline
all five models & $\Omega_{{\rm m}}=0.25\pm0.03$, $H_0=70.1\pm1.0$,
$w_0=-0.97^{+0.07, \, +0.19}_{-0.03, \, -0.20}$, $w_a=0.0^{+0.0, \,
  +0.8}_{-0.0, \, -0.8}$ \\ 
% & $a_p=0.79$, $w_p=-0.97^{+0.06, \, +0.14}_{-0.03, \, -0.10}$ \\
models I, II, \& V & $\Omega_{{\rm m}}=0.24\pm0.03$, $H_0=70.1\pm1.0$,
$w_0<-0.98, \, -0.86$, $w_a=0.0^{+0.0, \, +0.0}_{-0.0, \,
  -0.0}$\\ 
%& $a_p=0.96$, $w_p<-0.98, \, -0.86$ \\
\hline
\end{tabular}
\end{center}
\end{table*}

Alternatively, we can consider each parametrization as representing a
model, and within each parametrization marginalize over the different
choices of prior that we considered plausible. In this approach we
average the evidences (not their logarithms, as it is the evidences
themselves which represent the model probability) to obtain $\Delta
\ln E = -1.5$ for the constant $w$ model and $\Delta \ln E = -2.6$ for
the two-parameter model. The corresponding probabilities are then
77\%, 18\%, and 5\% for $\Lambda$CDM, the one-parameter, and
two-parameter dark energy models respectively. This approach gives
quite similar results to the above, while avoiding penalizing the
$\Lambda$CDM model for only having one choice of prior. However for
the remainder of the paper we will not average over models in this
way.

Finally, we might be interested only in a subset of the models; for
instance, we may consider only the models that do not allow $w<-1$
(models I, II, and V), motivated by quintessence models. Amongst these
models the probability is divided as 78\%, 21\%, and 1\% respectively.

Whichever the choice made, the overall conclusion is that the
$\Lambda$CDM model is preferred by present data, but that there are
non-negligible probabilities for the models of evolving dark
energy. We will explore the implications of this for future dark
energy searches in Section~\ref{secVB}.
 
\subsection{Parameter values and Bayesian model averaging}

We now consider the implications of the model selection framework for
constraints on the cosmological parameters. Within each model the
usual parameter probability distribution analysis applies, as given in
Section~\ref{s:params}. However we now need to combine these to
derive parameter constraints that account for model uncertainty (the
uncertainty in which model is the true model). The appropriate tool to
carry this out is \emph{Bayesian model averaging}, which is nicely
summarized by Hoeting et al.~\cite{hoeting}.\footnote{Bayesian model
averaging has only been used once previously in cosmology, in
interpretting simulated galaxy cluster data \cite{MHS}, and only very
occasionally in astrophysics/geophysics \cite{BMAastro}. A distinct
idea, closely related to the themes of this article, is Bayesian
survey design which averages an experimental figure of merit over a
set of possible cosmological models \cite{IPSO}.}

The basic idea is quite simple; rather than having a single
probability distribution for a parameter, we instead have a
superposition of its distributions in different models, weighted by
the relative model probability. In some models the parameter may have
a fixed value (e.g.~$w=-1$ in $\Lambda$CDM), and then that component
of the distribution is an appropriately-normalized delta-function. The
set-up is analogous to quantum mechanics; whilst the true model is
uncertain the distribution lies in a superposition of states, with the
possibility that future measurements may collapse the probability into
one of the models.

The posterior probabilities of the models are given via Bayes' theorem
by
\begin{equation}
P(M_k|D)=\frac{P(D|M_k)P(M_k)}{\sum_k P(D|M_k)P(M_k)}\,,
\end{equation}
where $P(D|M_k)$ is the evidence of model $M_k$. Here $P(M_k)$ are
prior model probabilities, which we take to be equal across the
models. Any other choice can be incorporated if required.

Within a gaussian approximation, it is easy to write down suitable
expressions for model averaging the parameter means and variances
\cite{hoeting}, but it is practically as easy to manipulate the full
distributions given by the parameter chains. One simply takes the
chains from each model and weights them according to the model
probability.  That the elements will have noninteger weights is no
problem (indeed CosmoNest chains, unlike those generated by Markov
Chain Monte Carlo, already have noninteger weights with the weights
within each chain summing to unity). All the chains can then be analyzed
together by the usual means such as the \texttt{getdist} package of
CosmoMC \cite{LB}.

\begin{figure*}
\centering
\includegraphics[width=14cm]{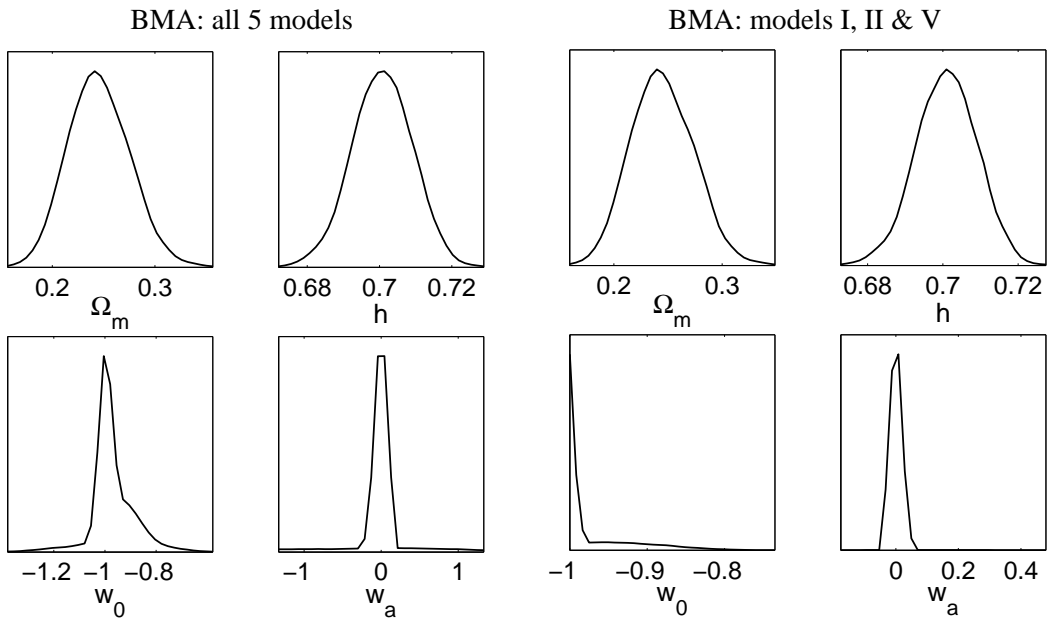}
\caption{Posterior parameter distributions obtained using the
WMAP+SDSS(BAO)+Astier05 data combination from Bayesian model averaging
(BMA). The left set of four panels averages over all the five models
under consideration, and the right over the quintessence-type models
(I, II and V) alone. Some smoothing of the delta-functions has been
carried out by binning.}
\label{fig3}
\label{fig4}
\end{figure*}

One shouldn't overstate the usefulness of this method, as the details
depend on a lot of prior information: the precise choice of models,
including their prior parameter ranges, to be averaged, and also the
prior model probabilities. Nevertheless there are some general
qualitative lessons to be learned.

The most important such lesson is that if one is seeking to limit a
parameter around some special fiducial value, eg $w=-1$ for the
equation of state, then the parameter errors are typically going to be
\emph{overestimated} if one ignores model uncertainty. The reason is
that in the absence of a detection, a substantial part of the model
probability is always going to be placed in the embedded model (in
this case $\Lambda$CDM), which adds a delta-function to the
probability distribution and hence suppresses the tails where the
limits will be imposed.

The parameter constraints obtained from Bayesian model averaging the
models together are summarized in Table~\ref{t2}. Because of the
presence of delta-functions in the averaged distribution, in some
cases confidence limits can be precisely zero.  The posterior
distributions for the parameters are shown in Fig.~\ref{fig3}, the
left set of panels showing averaging over all five models, and the
right set averaging the $\Lambda$ model with the quintessence-type
models (II and V).

The probability distributions of the parameters derived from current
data, after taking into account model uncertainty by Bayesian model
averaging over the models allowed by the data, summarize our current
state of knowledge regarding these parameters.  In this case we can
see that even though many models are still allowed by the data, given
the weight of the $\Lambda$CDM model, the constraints have tightened
significantly around $w_0=-1$ and $w_a=0$. For instance, compare the
model-averaged constraints of Fig.~\ref{fig3} (left) with the Model IV
constraints in Fig.~\ref{fig2}.

%For reference, the pivot parameters \cite{HT,DETF} derived from
%constraints on $w_0$ and $w_a$ are also given in the
%Table~\ref{t2}. To define the pivot parameters, one finds the scale
%factor $a_p$ at which $w(a)=w_0+w_a(1-a)$ is best constrained, given
%by $1-a_p=-\langle \delta w_0 \delta w_a \rangle/ \langle \delta w_a^2
%\rangle$. One then constrains the equation of state at that point,
%$w_p=w_0+w_a(1-a_p)$, along with $w_a$; by construction $w_p$ and
%$w_a$ are on average uncorrelated. For the second set
%of models, constraints on $w_p$ are very close to those on $w_0$
%because $a_p$ is so close to unity, with $w_a$ close to zero. 

Note that Bayesian model averaging is an intrinsic part of the model
selection framework, not an optional extra. As soon as one concedes
that there might be different model descriptions of data to which
probabilities should be assigned, consistent inference will then
require those probabilities to be properly accounted in deriving
parameter probability distributions. This is necessary too for
consistent model selection forecasting, as we now describe.

\section{Discussion: implications for future surveys}

\label{secVB}

We now consider the implications of these results for future surveys,
keeping this final discussion qualitative. In keeping with the
previous section, an analysis of future data should first assess the
validity of the various models being considered, and only then move on
to parameter estimation. The model comparison may be decisive, leaving
only one model on the table (which might be either $\Lambda$CDM or one
of the evolving models), or it may still leave several viable
models. If only one model survives then standard parameter estimation
tools will be valid, otherwise model averaging should again be
deployed to study parameter distributions.

The main aim of forecasting the power of future surveys is to enable
informed choices as to which projects to fund.  The Dark Energy Task
Force (DETF) recently produced an influential report \cite{DETF}
quantifying the capabilities of a wide range of proposed experiments
to constrain dark energy. Following ideas from Ref.~\cite{HT}, they
defined a Figure-of-Merit (FoM) as the inverse of the area inside the
95\% contour in the $w_0$--$w_a$ plane, for a fiducial $\Lambda$CDM
model. Normalizing to present knowledge, this factor is typically a
few to a few tens for proposed experiments of increasing
sophistication.\footnote{These ideas have also been extended beyond
survey comparison to the issue of survey design by Bassett and
collaborators \cite{IPSO}.}

The DETF FoM presumes that the two-parameter dark energy model is the
true one (i.e.~that $w_0$ and $w_a$ are parameters to be varied in
fitting the data), and quantifies the extent to which future
experiments will compress the allowed parameter range about the point
$w_0=-1$ and $w_a=0$. What it does not do is allow for the possibility
that the two-parameter model is not correct. To quote from the
abstract of Ref.~\cite{hoeting}, \textit{``Data analysts typically
select a model from some class of models and then proceed as if the
selected model had generated the data. This approach ignores the
uncertainty in model selection, leading to inferences that are more
risky than one thinks they are.''} One way to avoid this problem is to
employ model selection forecasting, as described in Ref.~\cite{M06},
which proposed a FoM based on the parameter area in which
$\Lambda$CDM cannot be strongly excluded using the Bayesian evidence.

We stress that the DETF FoM is a perfectly good way of distinguishing
the capabilities of different experiments, even though it is a
parameter estimation tool and those experiments are primarily seeking
to answer model selection questions. It is entirely reasonable to
believe that an experiment which is better at estimating parameters
within a model will also be better at model selection of that model
against embedded models. Our aim here is to advise caution against
over-interpretting the DETF FoM, in terms of the probability that an
upcoming experiment will actually detect dark energy evolution. 

The model selection considerations we have outlined have three
important implications in interpretting the DETF FoM.
\begin{enumerate}
\item \emph{The chances of detecting dark energy evolution are much
    less than implied by the fractional shrinkage of parameter area.}
    For example, if the FoM says the area in the $w_0$--$w_a$ plane
    will shrink by a factor 10, this does not mean a 90\% chance that
    evolution will be detected. This comment matches most people's
    intuition, but is quantified by the realization in model selection
    that a substantial part of the probability lies in the
    $\Lambda$CDM model. If this model is true, then obviously
    evolution cannot be detected as that would rule out the true
    model. Since present knowledge puts most of the probability in
    $\Lambda$CDM, as shown above, we can immediately conclude that the
    current chances of even an arbitrarily good experiment detecting
    dark energy evolution are less than half (with the significant
    caveat of the various model and parameter priors we have assumed).

\item \emph{There is a substantial probability that $\Lambda$CDM is
    the correct model, but the DETF FoM does not quantify how well
    experiments will determine this.} If $\Lambda$CDM is the true
    model, then the outcome of future experiments will be to support
    that model. This too would be a highly-valuable outcome. In this
    case, there is another model selection based FoM, described in
    Ref.~\cite{M06}, which evaluates the strength with which an
    upcoming experiment is expected to deliver a model selection
    verdict \emph{in favour} of $\Lambda$CDM under the assumption that
    that model is correct.  Model selection approaches have the
    crucial property that, unlike parameter estimation methods, they
    can accrue positive support for the simpler model. As shown in
    Ref.~\cite{M06}, advanced experiments are capable of decisively
    ruling out the two-parameter models in favour of $\Lambda$CDM (see
    also Ref.~\cite{Trotta06}). This then is the answer to the
    often-asked question, how far do we have to tighten constraints on
    dark energy parameters before we can start to believe that
    $\Lambda$CDM is the true model. This question is often asked with
    parameter estimation forecasting techniques in mind, but the
    answer lies in model selection. A design goal of future
    experiments should be that they are able to give a decisive
    verdict for $\Lambda$CDM if it is the true model.

\item \emph{If evolution is neither detected nor decisively excluded,
    the DETF FoM will \textbf{overestimate} the parameter errors.} It
    overestimates because it does not incorporate Bayesian model
    averaging. A powerful experiment that fails to detect evolution is
    bound to push most of the model probability into the $\Lambda$CDM
    model, so that the eventual combined parameter chain includes only
    a small fraction of elements from the $w_0$--$w_a$ model. That is
    to say, the delta-function of probability at $w=-1$ will contain
    most of the posterior distribution. So an experiment which does
    not detect evolution will impose more powerful constraints than
    the FoM indicates.
\end{enumerate}

None of the above affects the validity of the DETF FoM as a tool for
quantifying the capabilities of different experiments, though one
should bear in mind that it may prove inadequate if the true model is
\emph{more} complicated than the $w_0$--$w_a$ model \cite{Albrecht06}.
Nevertheless, while the DETF FoM may correctly rank experiments
relative to one another, since the principal goal of dark energy
experiments is one of model selection, we would advocate where
possible also analyzing their capabilities using model selection
forecasting tools as described in Ref.~\cite{M06} and this paper.

The approach we have outlined incorporates, modifies and extends the
Expected Posterior Odds (ExPO) technique pioneered by Trotta
\cite{Trotta05}.  This approach splits the model parameter space into
regions where different model selection verdicts are expected, and
then averages these over the current distribution in parameter space
to obtain a probability of each outcome. Of course, only by actually
doing the experiment do you discover which outcome does arise. Trotta
did not however fully implement the model selection/Bayesian model
averaging framework, as he computed the present probability
distribution within one model only, whereas in Ref.~\cite{pahud}
multiple models were included in an ExPO-type
forecast. Ref.~\cite{M06} extended ExPO to delineate parameter space
regions where different model selection outcomes would be expected,
and to define model selection figures of merit. The present paper
further extends the framework to estimation of parameter uncertainties
via Bayesian model averaging as well as calculation of model
probabilities.

\section{Conclusions}

We have carried out a model selection analysis of dark energy models,
updating and expanding on an earlier analysis by Saini et
al.~\cite{Saini04}. We find, as did they, that the preferred model is
the $\Lambda$CDM model, and indeed we find that the two-parameter
$w_0$--$w_a$ model is quite significantly disfavoured already by
present data.

We have made a first use of the concept of Bayesian model averaging
\cite{hoeting} to obtain current cosmological parameter
uncertainties. Bayesian model averaging generalizes the usual Bayesian
parameter estimation methods to the situation where the choice of
model is uncertain, and in the absence of detections typically
significantly \emph{strengthens} parameter constraints. Finally, we
have described how to use this framework to project the probabilities
of different outcomes to future dark energy experiments, and in
particular to interpret the meaning of the figure-of-merit introduced
by the Dark Energy Task Force \cite{DETF}.

We conclude that based on present knowledge the probability of future
experiments detecting dark energy evolution is rather small, unless
the various prior assumptions of our analysis prove to be
ill-founded. This is simply because present data places the majority
of the probability in the $\Lambda$CDM model. On the other hand,
high-precision experiments may be able to decisively support the
$\Lambda$CDM model, this ability being measured by a model selection
figure-of-merit given in Ref.~\cite{M06}. If $\Lambda$CDM is not
picked decisively, and neither is dark energy evolution detected, then
they can give tighter limits on dark energy parameters than one would
infer from the DETF figure-of-merit.

\begin{acknowledgments}
P.M., A.R.L., and D.P.\ were supported by PPARC (UK), and Y.W.\ in
part by NSF CAREER grant AST-0094335. A.R.L.\ thanks the Institute for
Astronomy, University of Hawaii, for hospitality while this work was
completed. We thank John Skilling for a series of useful discussions.
\end{acknowledgments}

\end{document}